\documentclass[aps,preprint]{revtex4-2}

\usepackage{color}
\usepackage{graphicx}
\usepackage{amsmath}
\begin{document}


\title{Stochastic periodic orbits in fast-slow systems with self-induced stochastic resonance}


\author{Jinjie Zhu}
\thanks{zhu.j.ag@m.titech.ac.jp}
\affiliation{Department of Systems and Control Engineering, Tokyo Institute of Technology, Tokyo 152-8552, Japan}
\affiliation{School of Mechanical Engineering, Nanjing University of Science and Technology, Nanjing 210094, China}
\author{Hiroya Nakao}
\affiliation{Department of Systems and Control Engineering, Tokyo Institute of Technology, Tokyo 152-8552, Japan}


\date{\today}

\begin{abstract}
Noise is ubiquitous in various systems. In systems with multiple timescales, noise can induce various coherent behaviors. Self-induced stochastic resonance (SISR) is a typical noise-induced phenomenon identified in such systems, wherein noise acting on the fast subsystem causes stochastic resonancelike boundary crossings. In this paper, we analyze the stochastic periodic orbits caused by SISR in fast-slow systems. By introducing the notion of the mean first passage velocity toward the boundary, a distance matching condition is established, through which the critical transition position of boundary crossing can be calculated. The theoretical stochastic periodic orbit can be accordingly obtained via gluing the dynamics along the slow manifolds. It is shown that the theoretical predictions are in excellent agreement with the results of Monte Carlo simulations for a piecewise linear FitzHugh-Nagumo system even for large noise. Furthermore, the proposed method is extended to the original FitzHugh-Nagumo system and also found to exhibit consistent accuracy. These results provide insights into the mechanisms of coherent behaviors in fast-slow systems and will shed light on the coherent behaviors in more complex systems and large networks.
\end{abstract}


\maketitle

\section{\label{sec:1}Introduction}
Nature is intrinsically noisy. While noise is often considered to be irrelevant to the realization of organized phenomena, noise does play constructive roles in many systems, e.g., gene expression \cite{Schmiedel2019}, climate dynamics \cite{Zheng2020}, neural spiking \cite{McDonnell2011}, etc. In particular, resonance induced by noise is one of the most intriguing phenomena and is not only of theoretical interest but also of practical importance. Among them, stochastic resonance (SR) \cite{Gammaitoni1998} attracts broad attention due to its counterintuitive behavior, which was first observed in the climate system \cite{Benzi1981}. In stochastic resonance, moderate intensity of noise can enhance the signal-to-noise ratio for periodically perturbed random systems. It has been extensively investigated both theoretically and experimentally in all kinds of disciplines \cite{Gammaitoni1998,Wagner2019,Wuehr2018,Yang2021}. In excitable systems, even without the external periodic signal, the coherence resonance (CR) \cite{Hu1993,Longtin1997,Pikovsky1997} manifests itself as coherent oscillations for intermediate noise. The mechanism of CR has been clarified by Pikovsky and Kurths via the first passage statistics \cite{Pikovsky1997}. Later, Lindner and Schimansky-Geier \cite{Lindner1999} considered an analytical approximation in the limit of large timescale separation and quantitatively described CR through the marginal probability density of the reduced system. They applied similar ideas to the piecewise linear FitzHugh-Nagumo system for both CR and SR \cite{Lindner2000}. More accurate results, such as the rate, coefficient of variation and diffusion coefficient, were obtained in the leaky integrate-and-fire model \cite{Lindner2002}. For detailed descriptions and quantitative approaches to the coherent behaviors in excitable systems, the readers are referred to Ref.\cite{Lindner2004}. On the other hand, Muratov {\it et al.} \cite{Muratov2005,LeeDeVille2005} found that noise on excitable systems with largely different timescales can provide distinct mechanisms of coherence. For noise acting on the fast variable, the resulting coherent oscillation was named self-induced stochastic resonance (SISR) \cite{Muratov2005,LeeDeVille2005}. SISR has been shown to account for the mixed-mode oscillations in a relaxation oscillator \cite{Muratov2008}. Besides, SISR may also explain the mechanism of anticoherence resonance for large noise \cite{Luccioli2006,Baspinar2020}. Recently, Yamakou {\it et al.} investigated SISR in multiplex neural networks and showed its potential application to optimal information processing \cite{Yamakou2019,Yamakou2020}.

Determining the oscillator's periodic orbit is important for capturing its dynamical behaviors under perturbations. For deterministic limit-cycle oscillators, the asymptotic phase can be correspondingly defined within the basin of attraction of the limit cycle and the phase reduction approach \cite{Kuramoto1984,Nakao2016} can be applied to reduce the dimensionality of the system for weak perturbations. However, it is not an easy task to determine the periodic orbit for stochastic oscillators. Some efforts have been made on the CR oscillator \cite{Zhu2020}. The noise-induced escape from the fixed point can be approximated by a jump process. The remaining part of the stochastic trajectory will be dominated by the noise-free system. Through this approximation, the phase reduction approach can be correspondingly established on this hybrid system \cite{Zhu2020}. Compared with the CR oscillator, the SISR phenomenon is more robust to parameter variations as the latter does not require the system to be close to bifurcation \cite{Muratov2005}. This gives SISR broader coherence ranges with respect to the noise strength. To depict the stochastic periodic orbit, Muratov {\it et al.} \cite{Muratov2005,LeeDeVille2005} applied the large deviation theory to find the critical transition position via the timescale matching condition. Recently, Yamakou and Jost \cite{Yamakou2018} further clarified the allowed interval for the potential difference and discussed the connection between inverse stochastic resonance and SISR. These analyses required both the slow timescale and the noise strength to approach zero. 

However, when the system has different timescales on different branches of slow manifolds as in the FitzHugh-Nagumo (FHN) model, there can be asymmetry in the noise-induced transition as shown later. In this study, to accurately obtain the stochastic periodic orbit in SISR oscillators, we define the mean first passage velocity and propose a condition that determines the critical transition position under the assumption that the transition process is continuous. This assumption differs from the previous studies, where the transition is considered as instantaneous and determined via the timescale matching condition \cite{Muratov2005,LeeDeVille2005}. We first employ this condition in a simplified piecewise linear FHN system and then extend it to the original FHN model. The theoretical results exhibit good consistency with the numerics even for large noise strength. Thus, the method to determine the periodic orbit proposed in this paper is robust and may have a wide range of applications.

The structure of this paper is as follows: In Sec.~\ref{sec:2}, the piecewise linear FitzHugh-Nagumo system is introduced and its dynamical feature is briefly discussed. Next, the self-induced stochastic resonance of this system is investigated in Sec.~\ref{sec:3} and the distance matching condition is proposed through which the critical transition positions on the left and right branches can be identified. Further, the theoretical stochastic periodic orbit is verified by Monte Carlo simulations. In Sec.~\ref{sec:4}, the proposed method is extended to the original FitzHugh-Nagumo system. Finally, conclusions and discussions are given in Sec.~\ref{sec:5}.

\section{\label{sec:2}Piecewise linear FitzHugh-Nagumo system}
\begin{figure}
	\includegraphics[width=0.6\textwidth]{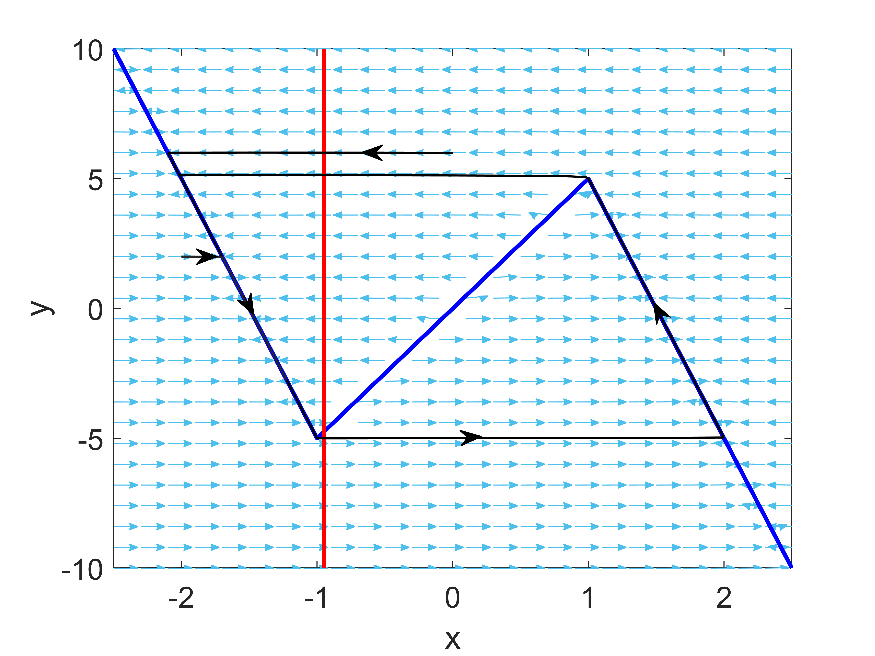}
	\caption{\label{fig:1}Vector field (cyan arrows) and deterministic trajectories (black lines) of system (\ref{eq:1}). The blue and red lines are $x$- and $y$-nullclines, respectively.}
\end{figure}

To illustrate the main idea of this paper, we choose the piecewise linear FitzHugh-Nagumo (PWL-FHN) system, also known as the McKean model \cite{McKeanJr1970}. This model preserves the essential characteristics of neuronal dynamics while assures computational simplicity. The governing equations are as follows:
\begin{equation}
	\begin{split}
		\dot x &= f_{\rm pwl}(x)-y,\\
		\dot{y} &= \varepsilon(x+a),
	\end{split}
	\label{eq:1}
\end{equation}
and
\begin{equation}
	f_{\rm pwl}(x)=
	\begin{cases} 
		-10x-15,  & \mbox{for } x<-1 \\
		5x, & \mbox{for } -1 \leq x \leq 1\\
		-10x+15, & \mbox{for } x>1
	\end{cases}
	\label{eq:2}
\end{equation}
where $x$ and $y$ represent the membrane potential and recovery variable, respectively. The timescale separation parameter $\varepsilon$ is assumed small, which is key to the realization of SISR. We set $\varepsilon=0.05$ and $a=0.95$ fixed in this paper. Since the bifurcation parameter $a<1$, system (\ref{eq:1}) has a stable limit cycle. This is different from the previous studies \cite{Muratov2005,LeeDeVille2005} in which the bifurcation parameter was taken in the excitable regime. Actually, whether the system is excitable or oscillatory is not important as the SISR phenomenon occurs away from the equilibrium or the deterministic limit cycle \cite{Muratov2005,LeeDeVille2005}. Considering the oscillatory situation allows SISR to be observed for small enough noise, which is helpful for the validation of the theoretical prediction (results for the excitable situation are also given in Appendix \ref{sec:A}, which are similar to those for the oscillatory situation discussed in the main text). The timescale separation can be observed in the vector field and the deterministic trajectories as shown in Fig.~\ref{fig:1}. The trajectory quickly converges to the left or right branches of the $x$-nullcline. The period of the limit cycle is dominated by the slow motion along them, which is determined by the slow subsystem $y$ in system (\ref{eq:1}). We denote the left, middle and right branches of the $x$-nullcline as $B_l$, $B_m$ and $B_r$, respectively. For fixed $y\in(-5,5)$, the fast subsystem $x$ has three equilibrium points, two of which on $B_l$ and $B_r$ are stable and the other on $B_m$ is unstable.
\begin{figure}
	\centering
	\includegraphics[width=0.48\textwidth]{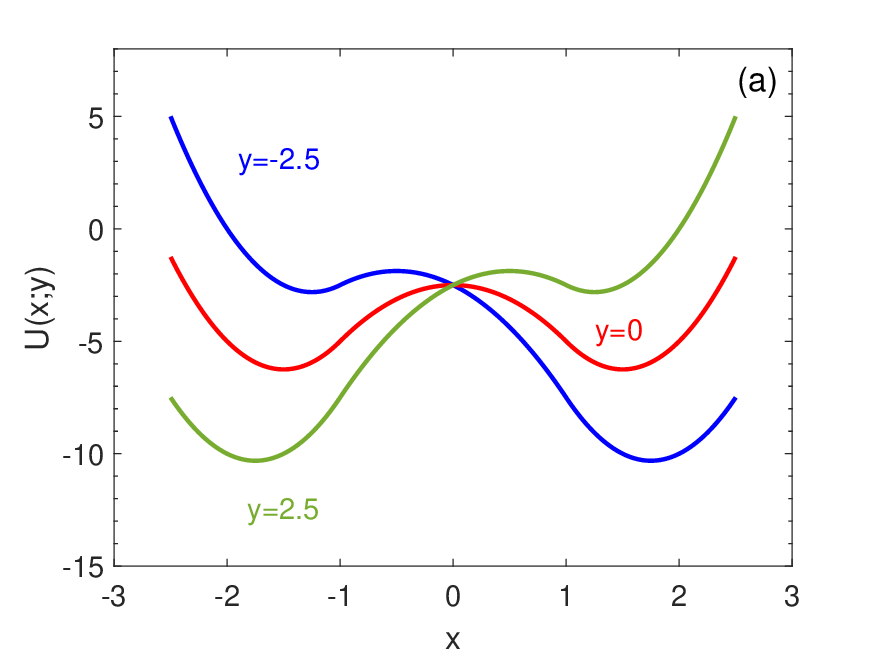}
	\includegraphics[width=0.48\textwidth]{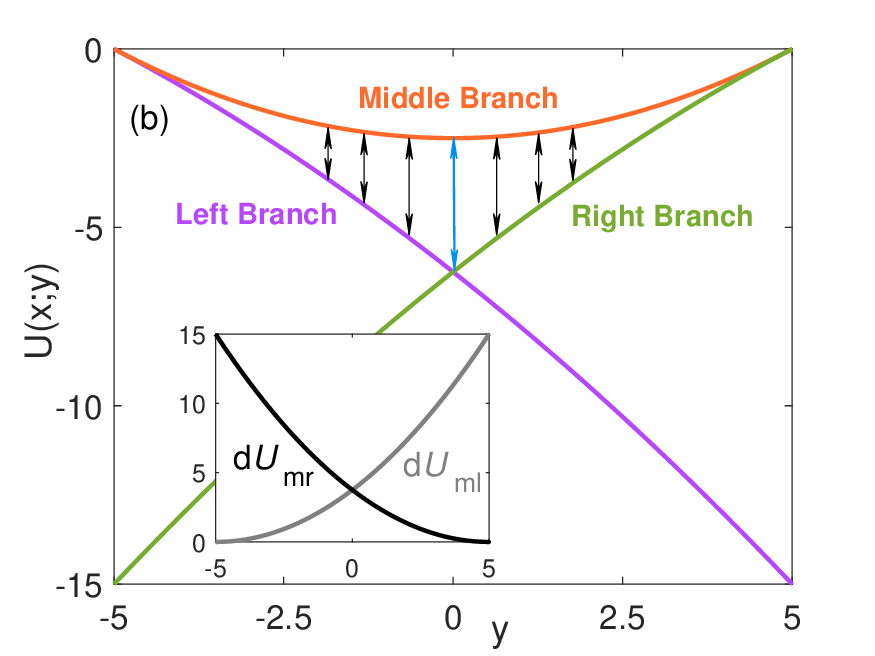}
	\caption{\label{fig:2}Potential function $U(x;y)$ for fixed $y$ values and on the three branches of the $x$-nullcline. (a) Potential function $U(x;y)$ for fixed $y$ values. (b) Potential values on the three branches of the $x$-nullcline of the fast subsystem in system (\ref{eq:1}) as functions of $y$ ($U_{\rm l}=-\frac{y^2}{20}-\frac{3y}{2}-\frac{25}{4}$, $U_{\rm m}=\frac{y^2}{10}-\frac{5}{2}$ and $U_{\rm r}=-\frac{y^2}{20}+\frac{3y}{2}-\frac{25}{4}$ denote the potential for the left, middle and right branches, respectively). The inset shows the potential differences between the middle branch and the left ($dU_{\rm ml}=U_{\rm m}-U_{\rm l}$) or right ($dU_{\rm mr}=U_{\rm m}-U_{\rm r}$) branch, respectively.}
\end{figure}

The potential value $U(x;y)$ of the fast subsystem $x$ of system (\ref{eq:1}) can be easily calculated as $U(x;y) =-\int f_{\rm pwl}(x)-y\, dx+C$ since it is piecewise linear, where $y$ is considered as a parameter and $C$ is a constant to be determined. Without loss of generality, we set the potential at the tips of the $x$-nullcline (i.e., $x=\pm1$) to be zero. Figure \ref{fig:2}(a)-(b) illustrate the potential $U(x;y)$ for fixed $y$ values and on the three branches of the $x$-nullcline, respectively. The potential difference between the middle branch and the left or right branch is monotonically decreasing with the system's evolution. They are symmetric to the line $y=0$ reflecting the symmetry of the $x$-nullcline.

\section{\label{sec:3}Stochastic periodic orbit of PWL-FHN system}
\begin{figure}
	\centering
	\includegraphics[width=0.6\textwidth]{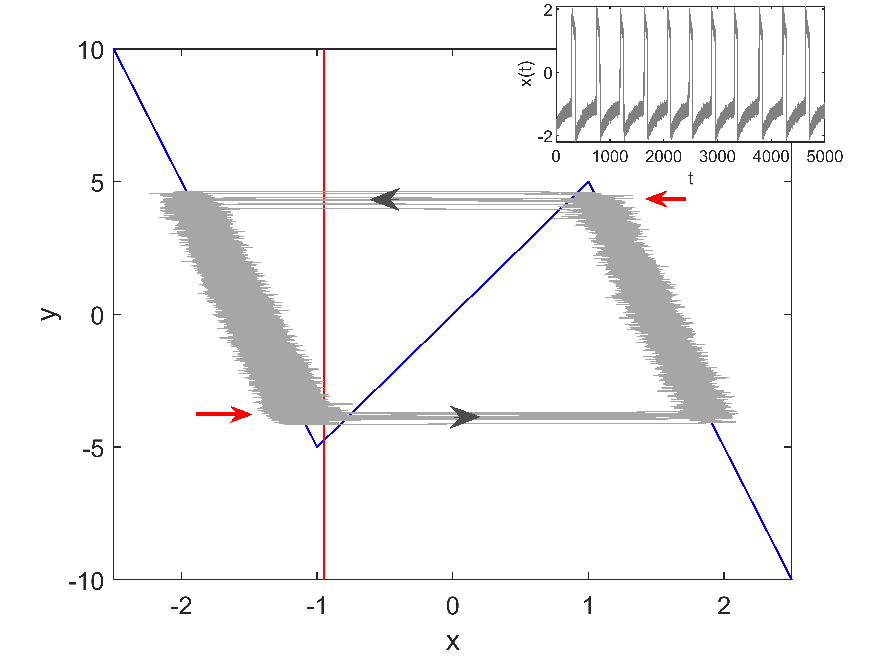}
	\caption{\label{fig:3}Self-induced stochastic resonance in PWL-FHN system. The gray line is the stochastic trajectory for $\sigma=0.1$. The inset shows the timeseries of the fast variable, which clearly exhibits coherent oscillations with an almost deterministic period. The red arrows indicate the asymmetry in the transition positions on the left and right branches.}
\end{figure}

As shown by Muratov {\it et al.} \cite{Muratov2005,LeeDeVille2005}, the SISR is observed when the noise acts on the fast variable. Therefore, we consider a stochastic version of the PWL-FHN system given as follows:
\begin{equation}
	\begin{split}
		\dot x &= f_{\rm pwl}(x)-y+\sqrt{\sigma}\xi(t),\\
		\dot{y} &= \varepsilon(x+a),
	\end{split}
	\label{eq:3}
\end{equation}
where $\xi(t)$ is Gaussian white noise with zero mean satisfying $\langle\xi(t)\rangle=0$ and $\langle\xi(t)\xi(\tau)\rangle=\delta(t-\tau)$. The parameter $\sigma$ determines the strength of the noise. Figure \ref{fig:3} illustrates the phase diagram and the timeseries for $\sigma=0.1$, which clearly shows the coherent behavior with a nearly deterministic period. It is notable that the stochastic periodic orbit caused by the noise is smaller than the deterministic limit cycle and that the orbit is not symmetric, namely, the transition to the other branch occurs earlier on the left branch than on the right branch. This asymmetry cannot be captured by the conventional timescale matching condition \cite{Muratov2005,LeeDeVille2005,Yamakou2018} as we will discuss below. The symmetric transition can only be realized for $a=0$, where the timescales on the two branches are the same owing to the symmetric vector field.

To determine the critical transition position on each branch, the conventional timescale matching condition assumes that the mean first passage time (MFPT) of boundary crossing and the recovery timescale matches with each other at the transition position. Following Muratov {\it et al.} \cite{Muratov2005}, the timescale matching condition is given by:
\begin{equation}
	{\rm exp}\left(\frac{\Delta W}{\sigma}\right) \sim \varepsilon^{-1},
	\label{eq:4}
\end{equation}
where $\Delta W$ stands for the quasipotential difference between the middle and the left or right branches. The quasipotential $W$ defined by Freidlin and Wentzell \cite{Freidlin2012} is similar to the potential function in the deterministic system, and the quasipotential difference $\Delta W$ characterizes the difficulty for the state to escape from a stable equilibrium. The left hand side (LHS) of Eq.(\ref{eq:4}) denotes the timescale of noise-induced transition of the fast variable while the right hand side (RHS) represents the timescale of recovery of the slow variable. By calculating the quasipotential difference of the fast subsystem by considering $y$ as a constant parameter, the critical position $y_*$ can be obtained, which predicts the transition position on each branch. Equation (\ref{eq:4}) is accurate in the vanishing noise limit and the slow relaxation limit, i.e., $\sigma \to 0$ and $\varepsilon \to 0$. Thus, according to this criterion, the critical positions would be symmetric with the same absolute value $|y_*|$ on the left and right branches. However, because $\varepsilon$ takes a finite non-zero value, this prediction is inconsistent with the observation as in Fig.~\ref{fig:3}.

To predict the noise-induced transition more accurately, we propose the following criterion. Using the MFPT $T_{\rm e}(y)$, we define the mean first passage velocity (MFPV) $V_{\rm e}(y)$:
\begin{equation}
	V_{\rm e}(y)=\frac{S(y)}{T_{\rm e}(y)},
	\label{eq:5}
\end{equation}
where $S(y)$ represents the distance of the left or right branch on which the state exits from the boundary (the middle branch). We expect that the transition occurs at $y_*$ at time $t(y_*)$ where the integration of the MFPV gives $S(y_*)$, i.e., 
\begin{equation}
	\int^{t(y_*)}_{t(y_0)} V_{\rm e}(y(t))\,dt=S(y_*),
	\label{eq:6}
\end{equation}
where $y_0$ is the starting position at time $t(y_0)$ of the trajectory on the branch under consideration. Equation (\ref{eq:6}) could serve as a more precise distance matching condition than Eq.(\ref{eq:4}) for solving the critical transition position $y_*$. This expression takes into account the slow variation in $y$ during the transition process; the LHS represents the approximate accumulation effect on the displacement of the fast variable $x$ as the slow variable $y$ moves from $y_0$ to $y_*$, and we assume that this accumulated displacement matches the distance $S(y_*)$ from the boundary at the transition position $y=y_*$.

We employ this idea for the PWL-FHN system. For simplicity, we consider the left branch only. The procedure for solving the critical position on the right branch is similar. The mean first passage time can be obtained from the well-known Kramers rate\cite{Kramers1940,Gardiner1985}:
\begin{equation}
	T_{\rm e}(y)=\frac{2 \pi}{\sqrt{\lvert U''_{\rm m}(x_y) \rvert U''_{\rm l}(x_y)}}{\rm exp}\left(\frac{\Delta W(y)}{\sigma}\right),
	\label{eq:7}
\end{equation}
where $U_{\rm m}(x_y)$ and $ U_{\rm l}(x_y)$ represent the potential function at the middle and left branches of the deterministic fast subsystem of system (\ref{eq:1}), respectively. The double prime denotes the second-order derivative with respect to the variable $x$, where the subscript $y$ indicates the $y$-position for the calculation of the potential function. Since system (\ref{eq:1}) is piecewise linear, $U''_{\rm m}(x_y)$ and $U''_{\rm l}(x_y)$ can be easily obtained as $-5$ and $10$, respectively. Because the fast subsystem is one-dimensional, the quasipotential difference $\Delta W=2\Delta U$, where $\Delta U=U_{\rm m}(x_y)-U_{\rm l}(x_y)=dU_{\rm ml}$ (see the inset in Fig.~\ref{fig:2}).

The distance can also be readily obtained by $S(y)=f_{\rm m}^{-1}(y)-f_{\rm l}^{-1}(y)$ for the left branch ($S(y)=f_{\rm m}^{-1}(y)-f_{\rm r}^{-1}(y)$ for the right branch), where the subscript indicates the specific branch (i.e., m, l and r for middle, left and right, respectively). Here, we have omitted the original subscript $\rm pwl$ for simple notification.

\begin{figure*}
	\centering
	\includegraphics[width=0.48\textwidth]{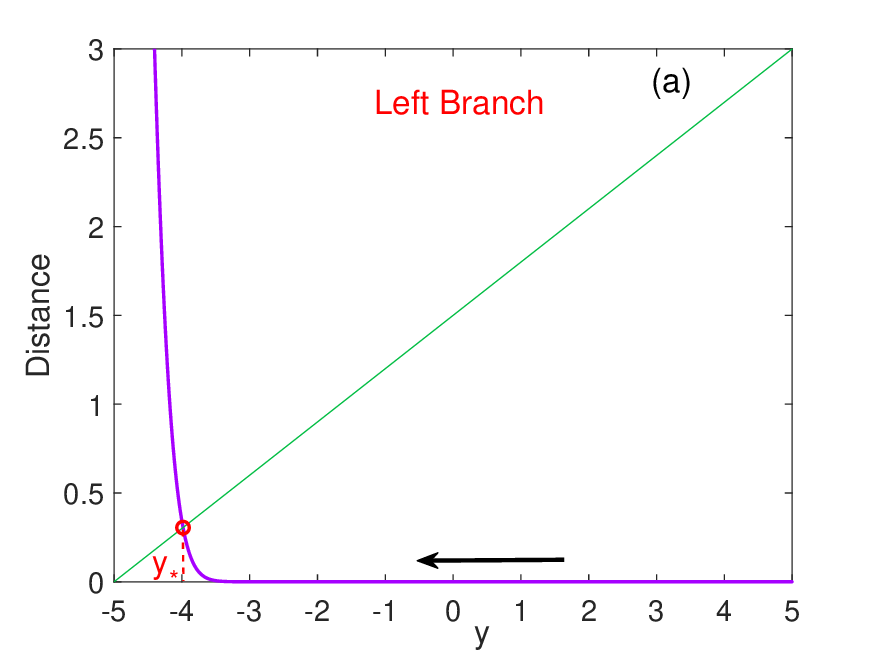}
	\includegraphics[width=0.48\textwidth]{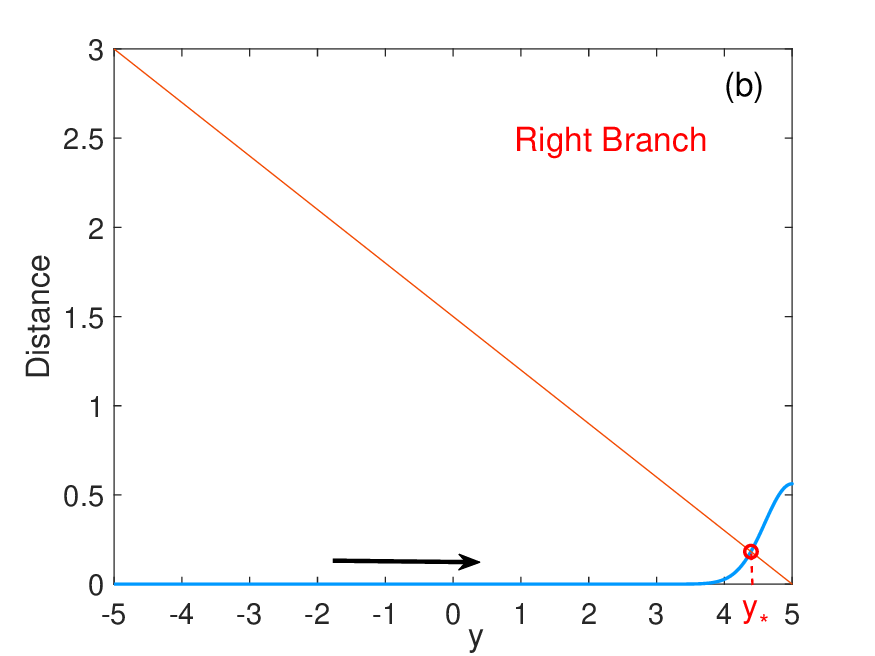}
	\caption{\label{fig:4}Identification of the critical transition position $y_*$ (intersection point). (a)Left branch. The thick  curve (purple) and the thin line (green) represent the LHS and RHS of Eq.(\ref{eq:8}), respectively; (b)Right branch. The thick curve (blue) and the thin line (orange) show the LHS and RHS of Eq.(\ref{eq:8}), respectively. The red circle denotes the intersection and the black arrow shows the direction of the system's evolution. The vertical coordinate represents the accumulated displacement of the fast variable $x$ towards the boundary (LHS of Eq.(\ref{eq:8})) and the distance between the middle and left or right branch (RHS of Eq.(\ref{eq:8})), respectively.}
\end{figure*}

By substituting the distance function $S(y)$ and the MFPT Eq.(\ref{eq:7}) into Eq.(\ref{eq:6}) and by using that $dy = \varepsilon( x + a ) dt$ and $x = f^{-1}(y)$, we can rewrite the condition for the critical transition position $y_*$ as
\begin{equation}
	\int^{y_{*}}_{y_{l}} \frac{V_{\rm e}(y)}{\varepsilon(f^{-1}_l(y)+a)}\,dy=\int^{y_{*}}_{y_{l}} \frac{S(y)\sqrt{\lvert U''_{\rm m}(x_y) \rvert U''_{\rm l}(x_y)}}{2 \pi \left(f^{-1}_l(y)+a\right)\varepsilon \, {\rm exp}\left(\frac{\Delta W(y)}{\sigma}\right)}\,dy=S(y_{*}),
	\label{eq:8}
\end{equation}
where we approximated $y_0$ by $y_l=5$ for the left branch. (A more appropriate, self-consistent choice of $y_0$ is the transition position on the other branch as obtained in Appendix \ref{sec:B} via the convergent method. However, the critical transition positions obtained by this method are nearly the same as those obtained by using the above two fixed values of $y_0$ as shown in Fig.~\ref{fig:13}. For simplicity, we fix $y_0=5$ and $y_0=-5$ on the left and right branches, respectively, in the rest of the main text.) By solving Eq.(\ref{eq:8}), we can obtain the critical transition position $y_*$. For illustration, we set $\sigma=0.1$. The values of the LHS and RHS of Eq.(\ref{eq:8}) are shown in Fig.~\ref{fig:4}(a) as functions of $y$. From the intersection of the two curves, the critical transition position on the left branch is evaluated as $y_* \approx -3.983$. Similarly, the critical transition position on the right branch is $y_* \approx 4.392$ (see Fig.~\ref{fig:4}(b)). 

It can be easily observed that the MFPV (thus the accumulated displacement in the direction of the fast variable $x$) remains nearly zero from the beginning ($y=y_l$) and quickly increases when the state approaches the transition position. This explains the mechanism of very coherent oscillations for SISR: before the transition position, the MFPV is nearly zero; after the transition position, the MFPV increases rapidly, which results in the noise-induced transition. The degree of coherence is mainly influenced by the timescale separation parameter $\varepsilon$. The smaller $\varepsilon$ is, the quicker the distance (LHS of Eq.(\ref{eq:8})) increases with $y$, and this leads to steeper curves in Fig.~\ref{fig:4}(a). Therefore, a small range of $y$ will determine the transition phenomenon. Furthermore, the left branch is closer to the $y$-nullcline than the right branch, which makes the evolution along the former slower than that along the latter. This causes the slower increase of LHS of Eq.(\ref{eq:8}) on the right branch, thus the transition on the right branch occurs later than on the left branch as shown in Fig.~\ref{fig:4} (this can also be quantitatively understood from the term $f^{-1}_l(y)+a$ in Eq.(\ref{eq:8})). In the limiting situation, i.e., for $\varepsilon \to 0$, as the other parts of the integrand in the LHS of Eq.(\ref{eq:8}) remains finite, $\varepsilon\,{\rm exp}\left(\frac{\Delta W}{\sigma}\right)$ should be of order $O(1)$, which is in fact the result obtained by Muratov {\it et al.} \cite{Muratov2005,LeeDeVille2005} as in Eq.(\ref{eq:4}).
\begin{figure}
	\centering
	\includegraphics[width=0.6\textwidth]{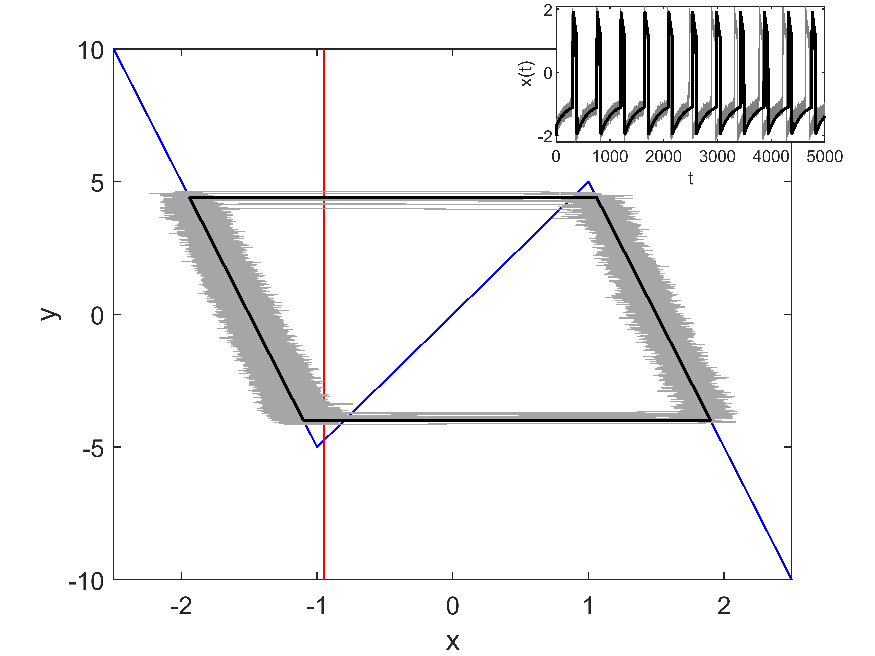}
	\caption{\label{fig:5}Stochastic periodic orbit predicted by Eq.(\ref{eq:8}) (bold black) compared with the stochastic trajectory (thin gray). Noise strength $\sigma=0.1$.}
\end{figure}
\begin{figure}
	\centering
	\includegraphics[width=0.48\textwidth]{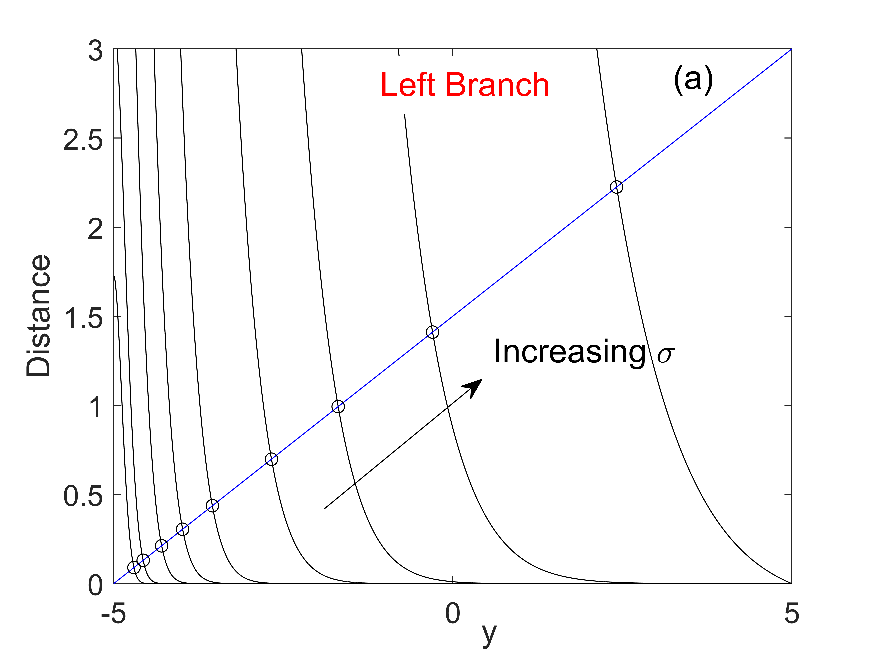}
	\includegraphics[width=0.48\textwidth]{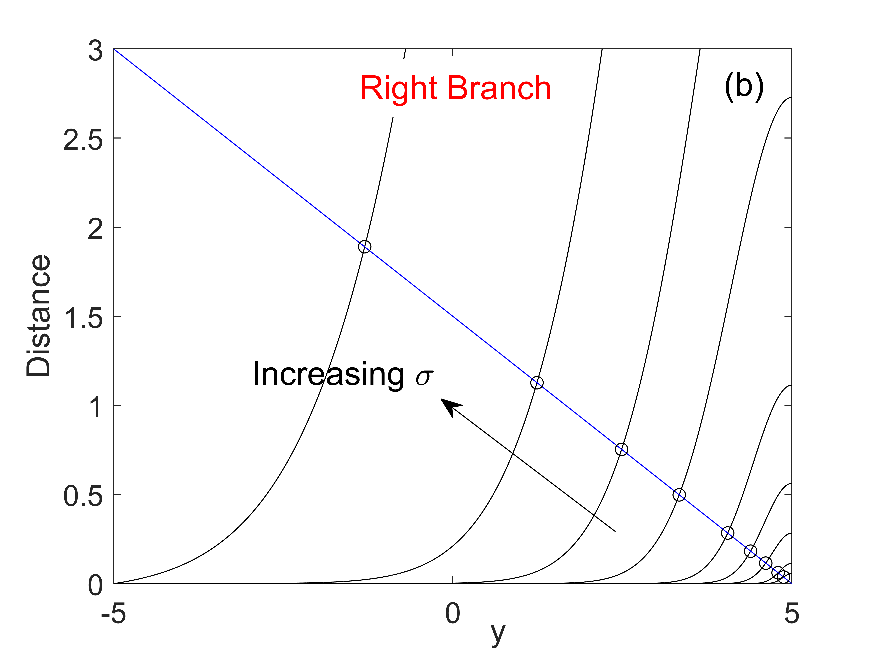}
	\caption{\label{fig:6}Identification of the critical transition position $y_*$ (intersection point). (a)Left branch; (b)right branch. The black curves and the blue line show the LHS and RHS of Eq.(\ref{eq:8}), respectively. The black circles denote the intersection points. Noise strengths are $\sigma=0.01, 0.02, 0.05, 0.1, 0.2, 0.5, 1, 2$ and $5$ (for the left branch, from left to right; for the right branch, from right to left).}
\end{figure}

Once we obtain the critical transition positions, the other parts of the stochastic periodic orbit can be determined by the deterministic slow dynamics along the left and right branches. Therefore, by connecting the transition positions, the total stochastic periodic orbit is obtained as shown in Fig.~\ref{fig:5}. It can be seen that the predicted stochastic periodic orbit is in excellent agreement with the stochastic trajectory obtained by the Monte Carlo simulation. Besides, it also predicts the asymmetric transitions on different branches. Following the same procedure, the critical transition position $y_*$ can be calculated for other noise strengths. We compute the results for $\sigma=0.01, 0.02, 0.05, 0.1, 0.2, 0.5, 1, 2$ and $5$, which are displayed in Fig.~\ref{fig:6}. The curves for the left branch are steeper compared with those on the right branch, which predicts earlier transition and higher coherence on the left branch. Note that, although we calculated the result for $\sigma=5$, it is practically impossible to achieve the corresponding transition positions since for $\sigma=5$ the critical position $y_*$ on the left branch is larger than that on the right branch.

\begin{figure}
	\centering
	\includegraphics[width=1\textwidth]{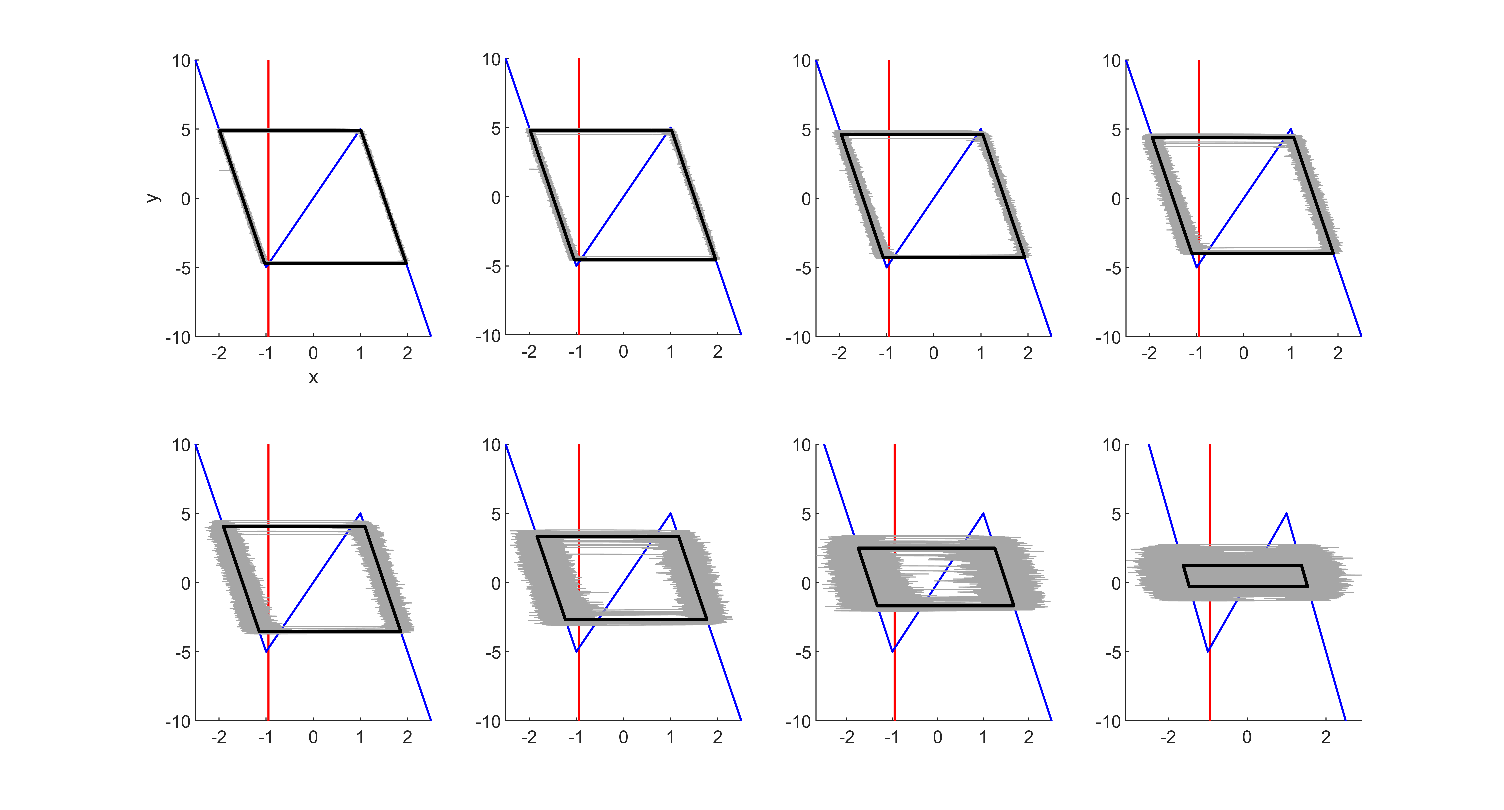}
	\caption{\label{fig:7}Stochastic periodic orbits predicted by Eq.(\ref{eq:8}) (bold black) and the stochastic trajectories (thin gray) for different noise strengths (from left to right, first row: $\sigma=0.01, 0.02, 0.05, 0.1$; second row: $\sigma=0.2, 0.5, 1, 2$).}
\end{figure}
To be more intuitive, the theoretical stochastic periodic orbits and the results of Monte Carlo simulations for various noise strengths are displayed in Fig.~\ref{fig:7} (similar results for the excitable PWL-FHN system are given in Appendix \ref{sec:A}). It can be seen that the theoretical predictions are in good agreement with the simulation results even for large noise, although the degree of coherence is deteriorated for increasing the noise strength. The stochastic periodic orbit becomes smaller as the noise strength $\sigma$ is increased, and at $\sigma_c \approx 2.733$, the critical transition positions on both branches become equal to each other. In this case, the stochastic periodic orbit will shrink into a line segment and the period will approach zero. The corresponding critical transition position is $y_{*c} \approx 0.51$, which is different from $y_{*c}=0$ in the case of the vanishing timescale separation ($\varepsilon \to 0$) \cite{Yamakou2018}. This prediction may account for the large-noise asynchrony in interacting excitable systems reported in Ref.~\cite{Touboul2020}.

\begin{figure}
	\centering
	\includegraphics[width=0.6\textwidth]{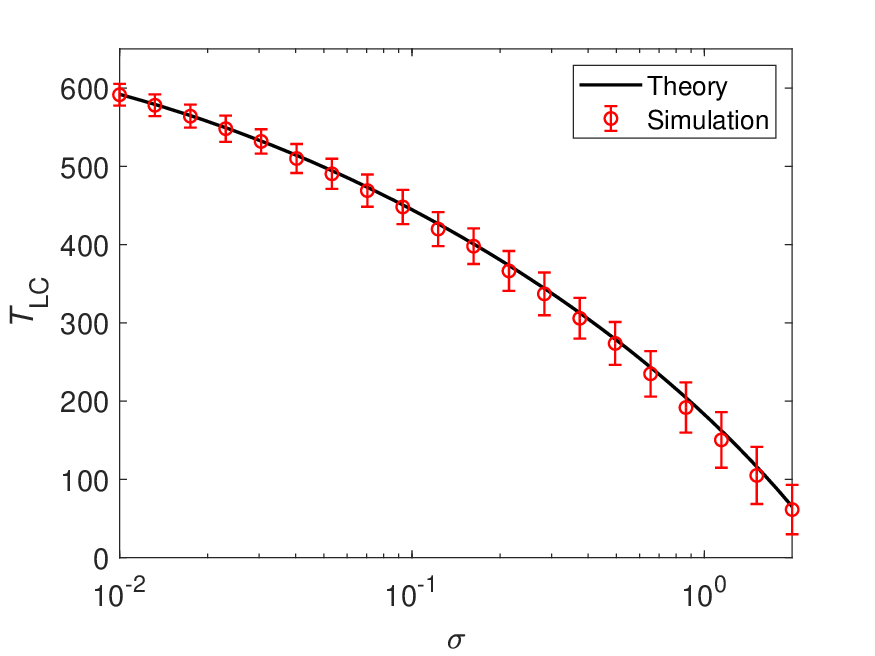}
	\caption{\label{fig:8}Period of stochastic periodic orbit versus noise strength. The black line denotes the theoretical prediction by Eq.(\ref{eq:9}), and the red circles are the results of Monte Carlo simulations, with the error bar showing the standard deviation.}
\end{figure}

The period of the stochastic oscillation, a fundamental quantity for its characterization, can also be predicted by using the proposed criterion. Because of the timescale separation, the period is dominated by the slow motion along the left and right branches. Therefore, the period of the stochastic periodic orbit can be approximated by the following equation:
\begin{equation}
	T_{\rm LC}=\int_{y_r}^{y_l} \frac{dy}{\varepsilon \left(f_l^{-1}(y)+a\right)}+\int_{y_l}^{y_r} \frac{dy}{\varepsilon \left(f_r^{-1}(y)+a\right)},
	\label{eq:9}
\end{equation}
where $y_l$ and $y_r$ represent the predicted critical transition positions on the left and right branches, respectively. The theoretical period of the stochastic periodic orbit for different noise strengths together with the results of Monte Carlo simulations are illustrated in Fig.~\ref{fig:8}. We can see that the prediction is in excellent consistency with the simulation results even for large noise. Besides, the small standard deviation demonstrates that the coherent oscillations persist over a broad range of noise strength, which is a significant feature of SISR that cannot be observed in CR \cite{Muratov2005}.

\section{\label{sec:4}Stochastic periodic orbit of FitzHugh-Nagumo system}
\begin{figure}
	\centering
	\includegraphics[width=1\textwidth]{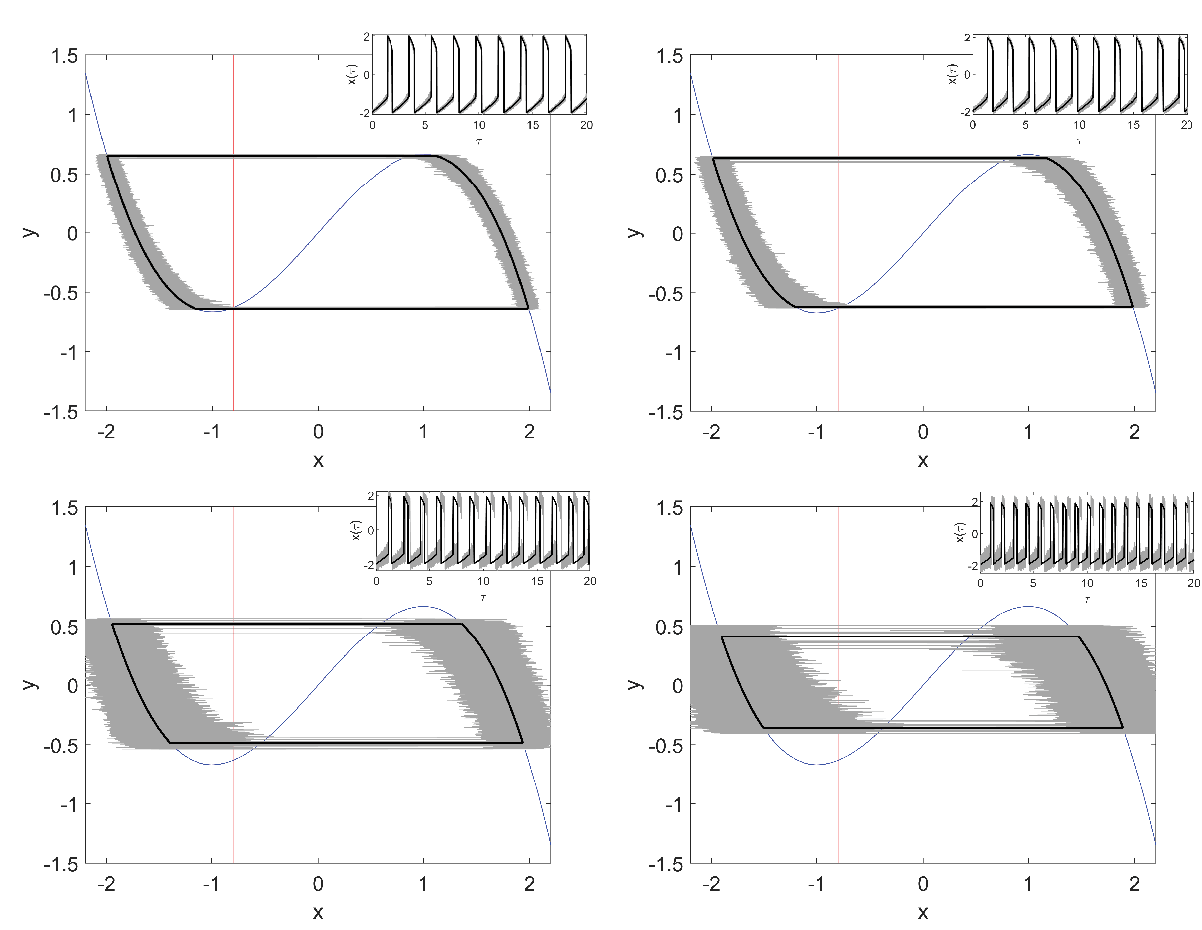}
	\caption{\label{fig:9}Stochastic periodic orbits of the FitzHugh-Nagumo system (\ref{eq:11}) predicted by Eq.(\ref{eq:8}) (black) and the stochastic trajectories obtained by the Monte Carlo simulations (gray) for different noise strengths (from left to right, first row: $\sigma=0.005, 0.01$; second row: $\sigma=0.05, 0.1$). The insets display the corresponding timeseries of the fast variable $x(\tau)$, where the slow time $\tau=\varepsilon t$ is used.}
\end{figure}

\begin{figure}
	\centering
	\includegraphics[width=0.6\textwidth]{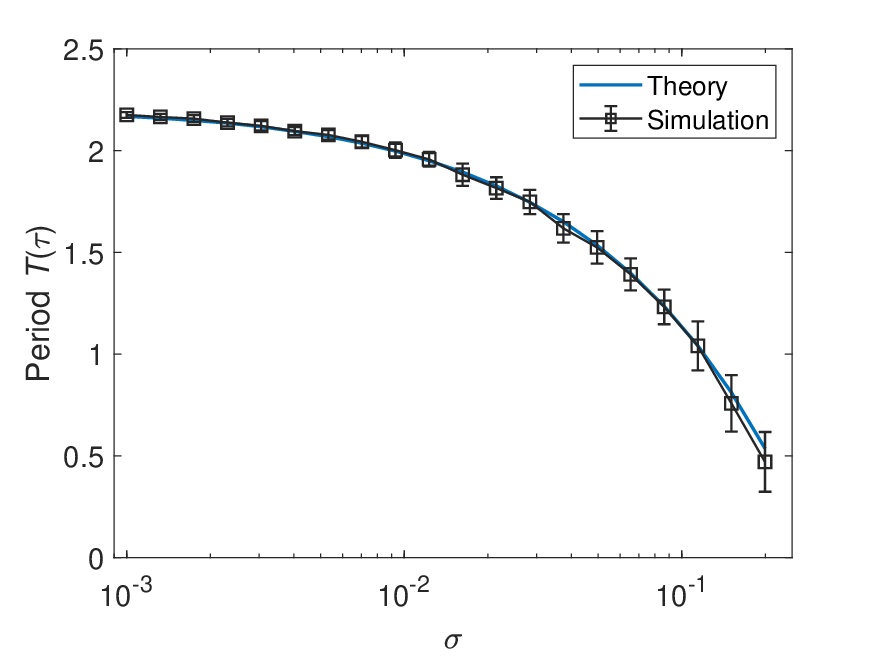}
	\caption{\label{fig:10}Period of stochastic periodic orbit versus noise strength for the FitzHugh-Nagumo system (\ref{eq:11}). The blue line denotes the theoretical prediction by Eq.(\ref{eq:9}), and the black squares are the results by the Monte Carlo simulations, with the error bar showing the standard deviation. The slow time $\tau=\varepsilon t$ is used.}
\end{figure}

To verify the validity of our theory in a more realistic model, we apply the proposed condition to the original FitzHugh-Nagumo system, i.e.,
\begin{equation}
	\begin{split}
		\dot x &= x-\frac{x^3}{3}-y+\sqrt{\sigma}\xi(t),\\
		\dot{y} &= \varepsilon(x+a),
	\end{split}
	\label{eq:10}
\end{equation}
where parameters are $\varepsilon=1{\rm e}-4$ as in Ref.~\cite{LeeDeVille2005} and $a=0.8$. The Gaussian white noise $\xi(t)$ is the same as that used in system (\ref{eq:3}). Following the previous procedure, the theoretical stochastic periodic orbits for different noise strengths can be numerically obtained via Eq.(\ref{eq:8}). They are illustrated in Fig.~\ref{fig:9}, which also shows excellent consistency with the results of Monte Carlo simulations. In the numerical simulations, to reduce the computational cost, we rescaled the equations by introducing a slow time $\tau=\varepsilon t$, which yields the noise strength multiplied with a coefficient $\sqrt{\varepsilon}$, i.e.,
\begin{equation}
	\begin{split}
		\varepsilon \frac{{\rm d}x}{{\rm d}\tau} &= x-\frac{x^3}{3}-y+\sqrt{\varepsilon\sigma}\xi(\tau),\\
		\frac{{\rm d}y}{{\rm d}\tau} &= x+a.
	\end{split}
	\label{eq:11}
\end{equation}
In a similar way to the PWL-FHN system, the period of the stochastic periodic orbit of the FHN system can be calculated as in Fig.~\ref{fig:10}. Again, the theoretically predicted period of the stochastic orbit in the FitzHugh-Nagumo system (\ref{eq:11}) shows excellent agreement with the results of Monte Carlo simulations.

\section{\label{sec:5}Conclusions and discussions}
In summary, we have investigated the stochastic periodic orbit in the SISR oscillator. By introducing the notion of the mean first passage velocity, the distance matching condition is established, through which the critical transition position on each branch can be determined. The theoretical stochastic periodic orbit is obtained by gluing the motions along the slow manifolds. By comparison with the results of Monte Carlo simulations, the theoretical predictions of the critical transition positions and periods of stochastic oscillations are proved to have good accuracy even for large noise. One of the main differences between the timescale matching condition by Muratov {\it et al.} \cite{Muratov2005} and our proposed condition is that the transition in the former is assumed to be instantaneous while it is considered to be continuous in the latter. As observed in Figs.~\ref{fig:7} and \ref{fig:9}, the fluctuations of the stochastic trajectories along $x$-axis gradually increase as they approach the transition point. This is due to slow change in the slow variable during the transition process and our condition approximately incorporates this effect.

From the derived distance matching condition Eq.(\ref{eq:8}), we can see that the timescale separation parameter $\varepsilon$ is significant to the observation of the SISR phenomenon. Because it controls the slope of the integration curve given by the LHS in Eq.(\ref{eq:8}), the smaller $\varepsilon$ corresponds to the larger slope, thus induces more coherent stochastic oscillations. This mechanism is briefly discussed in Appendix \ref{sec:C}. However, for the effect of noise strength on the degree of coherence, the slope of the integration curve does not give similar conclusions. This can be noticed by comparing the results in Fig.~\ref{fig:6} and Fig.~\ref{fig:8}, where larger slopes don't imply higher coherence. This is because increasing noise strength will not only increase the slope of the integration curve (this can be inferred from Eq.(\ref{eq:8}), similar to $\varepsilon$), but also increase the amount of fluctuations. The competition between these two impacts determines the effect of noise strength on the degree of coherence.

The definition of the mean first passage velocity in Eq.(\ref{eq:5}) is a natural extension of the mean velocity in the deterministic system. Instead of Eq.(\ref{eq:8}), we may also consider different but similar conditions. For example, the distance function $S(y)$ can be replaced by a constant 1 in Eq.(\ref{eq:8}), then the mean first passage velocity will be replaced by the mean passage frequency and the corresponding matching condition will give the critical position for the first transition. Another possibility is to replace $S(y)$ by the quasipotential difference $\Delta W(y)$, though the physical meaning may be not so intuitive. We have tested these two matching conditions and obtained similar predictions (although not as good as the distance matching condition in Eq.(\ref{eq:8}) as shown in Appendix \ref{sec:D}, wherein the comparison with the results via the timescale matching condition (\ref{eq:4}) is also displayed).

For higher dimensional systems, the computation of the quasipotential would be a major obstacle. Recent advances in the numerical methods for solving the quasipotential \cite{Lin2020,Kikuchi2020} would offer the possibility for the application of the proposed method to more complex situations, such as the prediction of inverse stochastic resonance \cite{Zhu2021}.

\begin{acknowledgments}
	We thank the anonymous reviewers for valuable and insightful comments, which have improved our manuscript substantially. J.Z. acknowledges the support from JSPS KAKENHI JP20F40017, Natural Science Foundation of Jiangsu Province of China (BK20190435) and the Fundamental Research Funds for the Central Universities (No.30920021112). H.N. thanks JSPS KAKENHI JP17H03279, JP18H03287, JPJSBP120202201, and JST CREST JP-MJCR1913 for financial support.
\end{acknowledgments}

\appendix
\section{\label{sec:A}Stochastic periodic orbits of the excitable PWL-FHN system}
\begin{figure}
	\centering
	\includegraphics[width=1\textwidth]{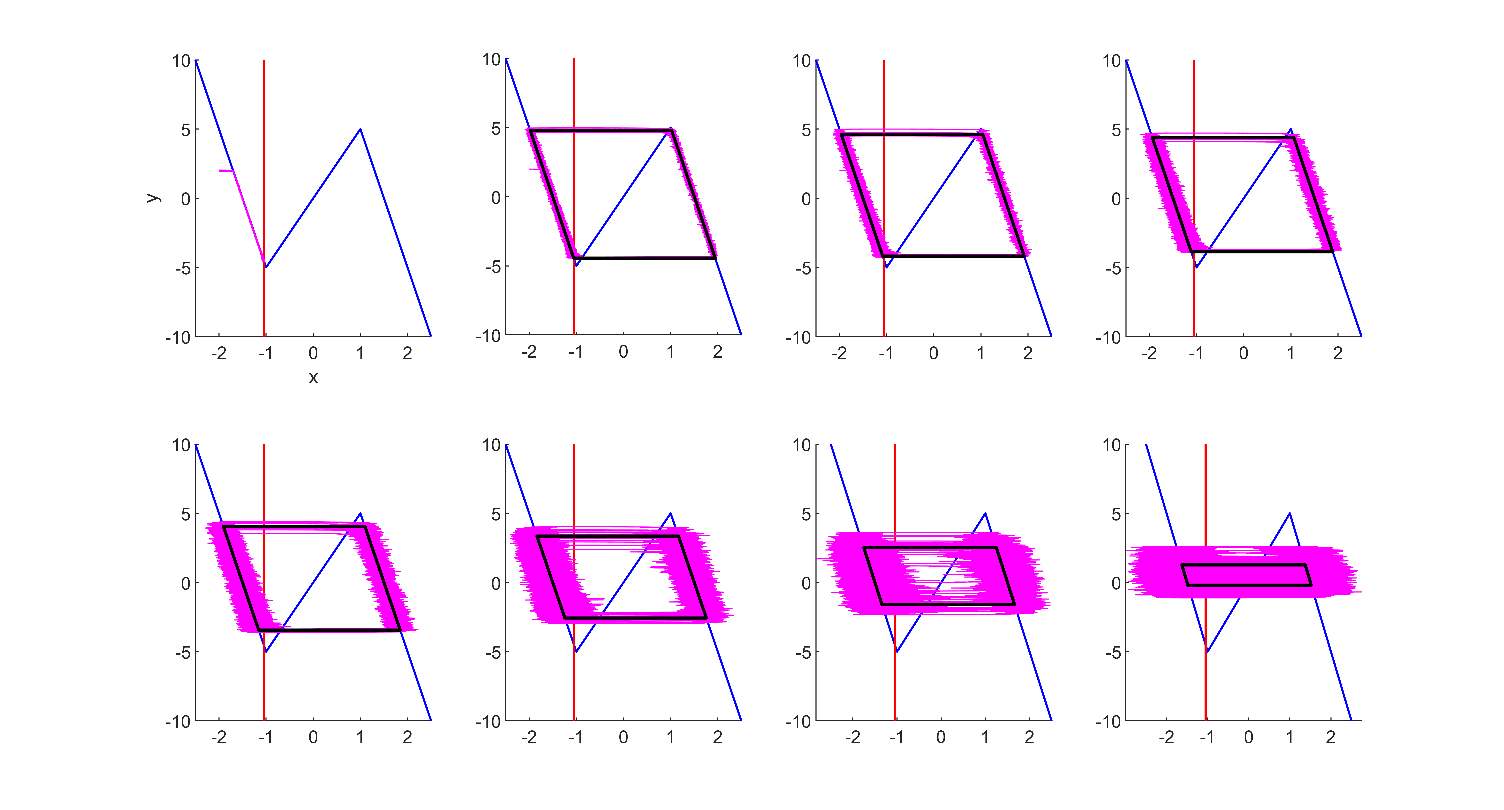}
	\caption{\label{fig:11}Stochastic periodic orbits predicted by Eq.(\ref{eq:8}) (bold black) and the stochastic trajectories (thin purple) for the excitable PWL-FHN system ($a=1.05$) with different noise strengths (from left to right, first row: $\sigma=0.001, 0.02, 0.05, 0.1$; second row: $\sigma=0.2, 0.5, 1, 2$).}
\end{figure}

In the main text, it is explained that whether the system is excitable or oscillatory is not important for our prediction. However, since the original work by Muratov {\it et al.}\cite{Muratov2005} considered the excitable case, it is interesting to see that the proposed distance matching condition works also for this situation. We consider the PWL-FHN system (\ref{eq:3}) and set the bifurcation parameter at $a=1.05$ while keeping the other parameters unchanged as in the main text. The stochastic periodic orbits predicted by utilizing the distance matching condition (\ref{eq:8}) are shown in Fig.~\ref{fig:11}, which are also in good agreement with the Monte Carlo simulation results.

\begin{figure}
	\centering
	\includegraphics[width=0.6\textwidth]{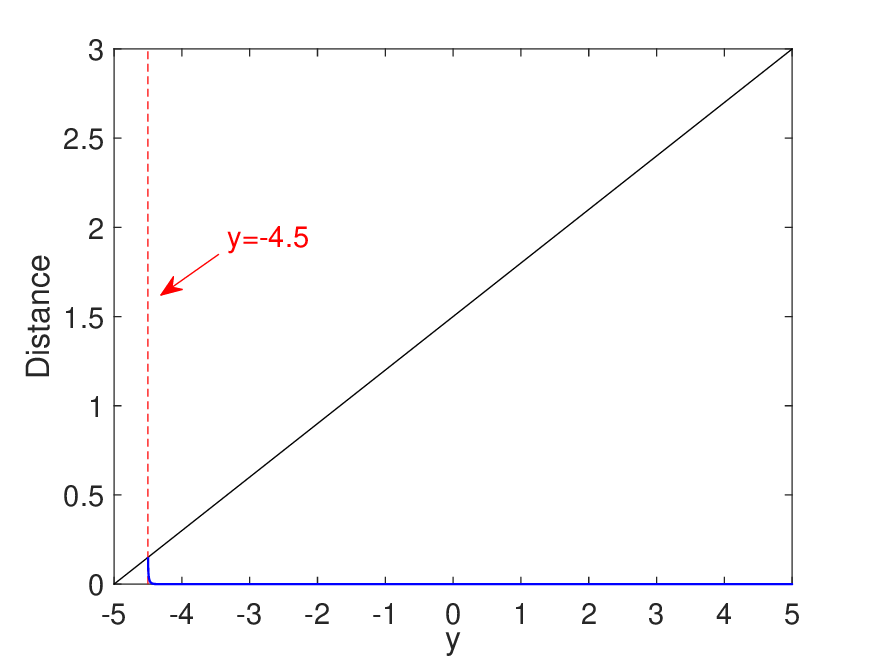}
	\caption{\label{fig:12}Illustration of the lower bound of noise strength ($\sigma \approx 0.01$) for the excitable PWL-FHN system ($a=1.05$). The red dashed line indicates the position of the equilibrium. The blue curve and the black line show the LHS and RHS of Eq.(\ref{eq:8}), respectively. Below this noise strength, the LHS and RHS of Eq.(\ref{eq:8}) cannot intersect before the system state reaches the equilibrium. The system is attracted by the equilibrium before a transition to the other branch can occur. Consequently, spike initiation becomes rare and the resulting stochastic oscillations are no longer coherent.}
\end{figure}

Unlike the oscillatory case, for the excitable system, when noise is small enough, the state will converge to the equilibrium before the transition. There is a lower bound of noise strength($\sigma \approx 0.01$, see Fig.~\ref{fig:12}), below which the stochastic orbit is no longer coherent or even difficult to initiate spikes(see Fig.~\ref{fig:11} for $\sigma=0.001$).

\section{\label{sec:B}Convergent method for critical transition positions}
For computing the critical transition position using Eq.(\ref{eq:8}), we approximated the starting position $y_0$ by $y_l=5$ for the left branch. However, a more accurate and self-consistent choice of $y_0$ would be the critical transition position on the right branch. Similarly, for the right branch, a more accurate choice would be the critical transition position on the left branch. We here calculate these self-consistent transition points and compare the results with our approximation.

To obtain the final critical transition positions on both branches, we can apply a convergent method as follows. First, we start at $y_0=5$ on the left branch and obtain the candidate transition position $y_{*l}$. We utilize this value as the starting position $y_0$ on the right branch and calculate the candidate transition position on the right branch $y_{*r}$. Then, $y_{*r}$ will be chosen as the starting position on the left branch to compute the next candidate transition position. We continue this process until the difference between adjacent candidate transition positions on each branch is less than a tolerance value (e.g., 0.001).

\begin{figure}
	\centering
	\includegraphics[width=0.6\textwidth]{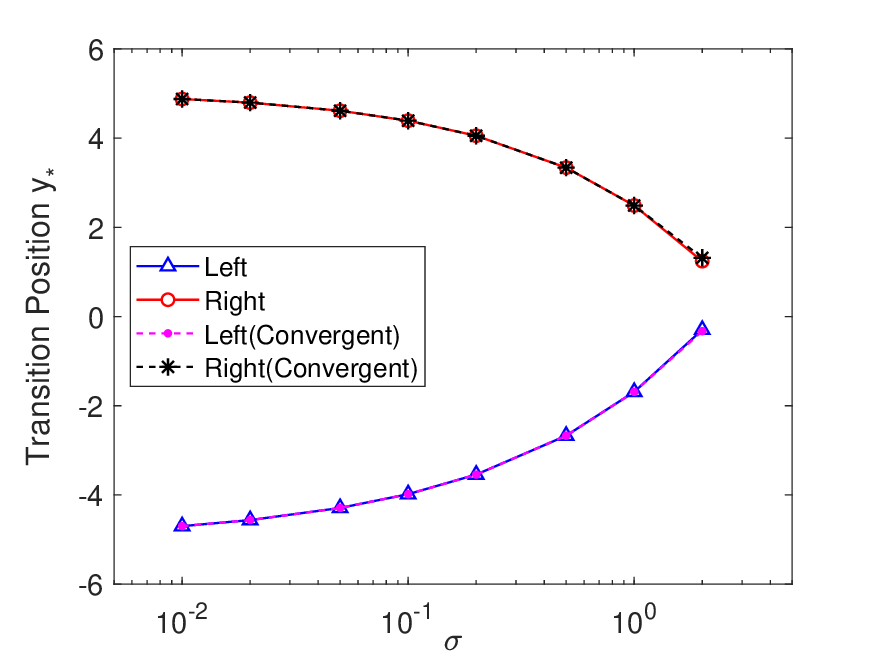}
	\caption{\label{fig:13}Comparison between the critical transition positions on left and right branches for fixed starting positions (solid) ($y_0=5$ on the left branch and $y_0=-5$ on the right branch) and those obtained by the convergent method (dashed). Parameters are the same as in Fig.~\ref{fig:6}.}
\end{figure}

The results are shown in Fig.~\ref{fig:13}. It can be seen that the critical transition positions obtained by fixed starting positions ($y_0=5$ on the left branch and $y_0=-5$ on the right branch) almost coincide with those obtained by the convergent method explained above. Only negligibly small differences can be observed when the transition positions on both branches are close enough. This result can be similarly explained as in the main text that only a small range of $y$ interval contributes to the integration of the LHS of Eq.(\ref{eq:8}). In general, the fixed starting positions used in the main text are enough to assure the accuracy of the transition position.

\section{\label{sec:C}Influence of the timescale separation parameter on the degree of coherence}
\begin{figure}
	\centering
	\includegraphics[width=0.6\textwidth]{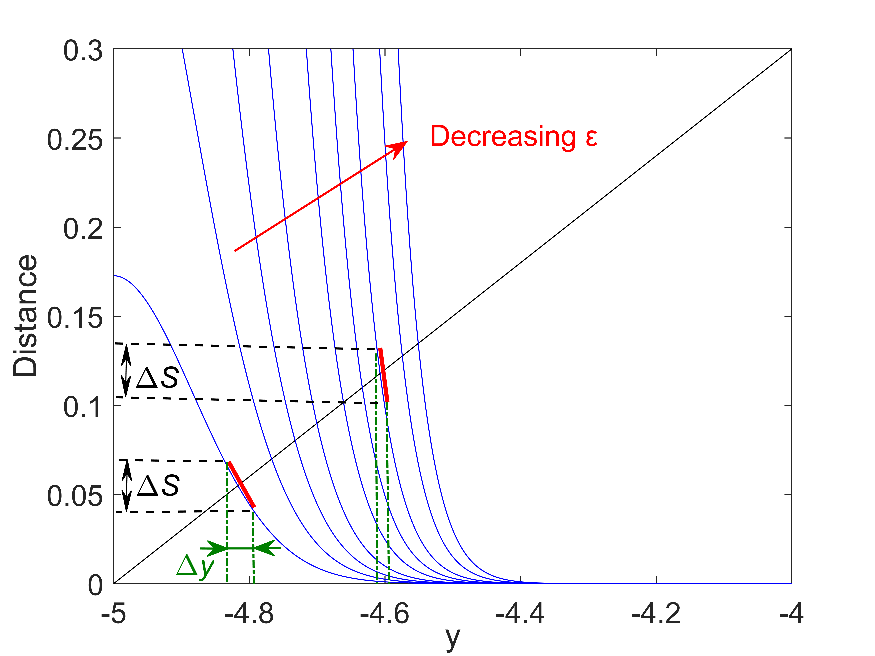}
	\caption{\label{fig:14}Illustration of the influence of the timescale separation parameter on the degree of coherence for the PWL-FHN system (\ref{eq:3}). The blue curves (from left to right, $\varepsilon=0.5, 0.2, 0.1, 0.05, 0.02, 0.01, 0.005, 0.002, 0.001$) and the black line show the LHS and RHS of Eq.(\ref{eq:8}) (left branch), respectively. The noise strength is $\sigma=0.01$.}
\end{figure}

It is intuitive that decreasing the timescale separation parameter $\varepsilon$ increases the degree of coherence of the SISR oscillator. Figure \ref{fig:14} illustrates the influence of $\varepsilon$ on the integration curve, i.e., LHS of Eq.(\ref{eq:8}) of the PWL-FHN system (\ref{eq:3}) for the left branch. It is shown that smaller $\varepsilon$ corresponds to a larger slope (absolute value) at the intersection point. For the same range of distance $\Delta S$, a larger slope implies a smaller range $\Delta y$. This means that a smaller range of $y$ determines the transition process, therefore the stochastic orbits exhibit higher coherence. Similar results also hold for the right branch of Eq.(\ref{eq:8}). Note that this mechanism is for fixed noise strength. The influence of noise strength on the degree of coherence is complex as explained in the main text.

\section{\label{sec:D}Comparison between different methods on the prediction of stochastic periodic orbits}
\begin{figure}
	\centering
	\includegraphics[width=0.6\textwidth]{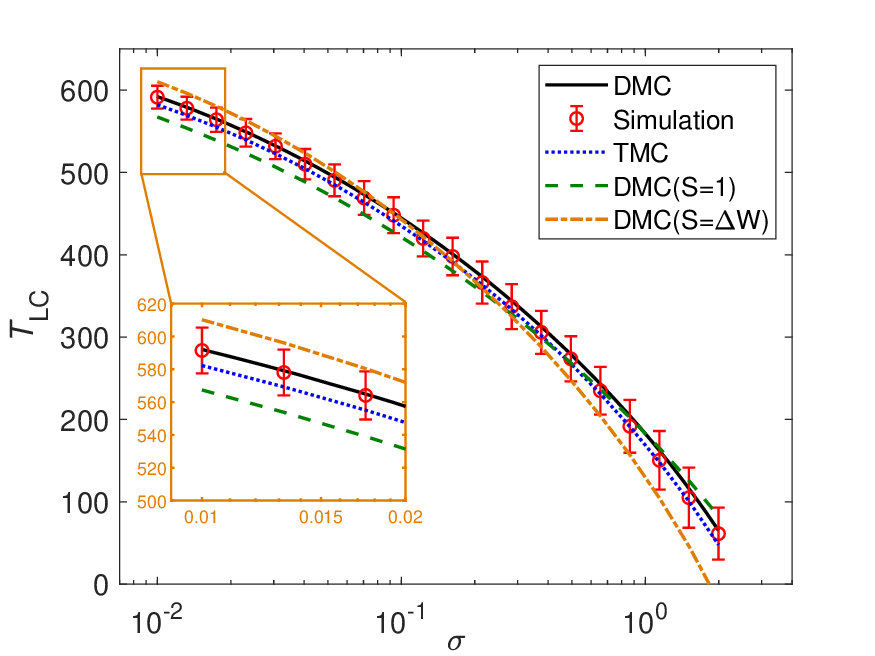}
	\caption{\label{fig:15}Period of stochastic periodic orbit of the PWL-FHN system (\ref{eq:3}) via different methods. The black solid curve and the red circles with error bars are the same as in Fig.~\ref{fig:8}. DMC (black solid) stands for the proposed distance matching condition Eq.(\ref{eq:8}). TMC (blue dotted) represents the timescale matching condition by Muratov {\it et al.} \cite{Muratov2005}. The green dashed and orange dot-dashed curves show the results by replacing the distance function $S(y)$ in Eq.(\ref{eq:8}) by constant 1 and quasipotential difference $\Delta W$, respectively. The inset displays the local magnification. Other parameters are the same as in Fig.~\ref{fig:8}.}
\end{figure}

As is discussed in Sec.~\ref{sec:5}, there are two other replacements of the distance function $S(y)$ in the distance matching condition Eq.(\ref{eq:8}), i.e., the constant 1 and the quasipotential difference $\Delta W(y)$. We test them in the PWL-FHN system (\ref{eq:3}) by computing the period of the stochastic orbits as shown in Fig.~\ref{fig:15}. For comparison, we also apply the timescale matching condition (\ref{eq:4}) by Muratov {\it et al.} \cite{Muratov2005}, where the critical transition positions on the left and right branches can be easily obtained as $y_{*}=\pm (-5+\frac{\sqrt{30}}{3} \sqrt{\sigma \ln(\varepsilon^{-1})})$. It can be seen that the distance matching condition used in the main text is more accurate than the others.



\end{document}